\documentclass{aa} 

\usepackage{graphicx}
\usepackage{longtable,lscape}
\usepackage{natbib}
\bibpunct{(}{)}{;}{a}{}{,}
\usepackage[usenames]{color}

\usepackage{txfonts}
\begin{document}
 \title{Periodicity search as a tool for disentangling the contaminated 
colour light curve
of CoRoT 102781750\thanks{
The CoRoT space mission was developed and is operated by the French space agency 
CNES, with participation of ESA's RSSD and Science Programmes, Austria, Belgium, Brazil, 
Germany, and Spain. Follow-up observations were obtained at Piszk\'estet\H{o}, the mountain 
station of Konkoly Observatory. 
}}

\author{M.~Papar\'o\inst{1}\fnmsep\thanks{\email{paparo@konkoly.hu}}
         \and
              M.~Chadid\inst{2}
         \and
              E.~Chapellier\inst{2}
         \and
              J.~M.~Benk\H{o}\inst{1}
         \and
              R.~Szab\'o\inst{1}
         \and
              K.~Kolenberg\inst{3}
         \and
              E.~Guggenberger\inst{3}
         \and
              Zs.~Reg\'aly\inst{1}	      
 	\and
              M.~Auvergne\inst{4}
	\and
              A.~Baglin\inst{4}
	\and
              W.~W.~Weiss \inst{3}
}

  \offprints{M.~Papar\'o}

  \institute{
    Konkoly Observatory of the Hungarian Academy of Sciences, Konkoly-Thege M. u 15-17, H-1121 Budapest, Hungary
\and
    Observatoire de la C\^ote d'Azur, Universit\'e Nice Sophia-Antipolis, UMR 6525, Parc
    Valrose, 06108 Nice Cedex 02, France
\and
    Institute of Astronomy, University of Vienna, T\"urkenschanzstrasse 17, A-1180 Vienna, Austria
\and
     LESIA, Universit\'e Pierre et Marie Curie, Universit\'e Denis Diderot, Observatoire de
     Paris, 92195 Meudon Cedex, France
}

\date{Received; accepted }

\abstract
 {
The CoRoT space mission (COnvection, ROtation and planetary Transits) launched in December 2006, 
aims at finding transiting exoplanets and investigating stellar oscillation in adjacent stellar 
fields, called exo- and seismofields, respectively.  Besides the seismofields, CoRoT has a 
strong potential for seismological research on the exofields. Up to now, only a limited 
number of RR\,Lyrae stars have been classified among the CoRoT targets. Knowing the astrophysical
importance of the RR\,Lyrae stars, we attempted to get useful information even
from a contaminated light curve of a possible RR\,Lyrae pulsator.
}
 {
The star CoRoT\,102781750 
reveals a puzzle, showing a very complex and altering variation in different 
`CoRoT colours'.  We established without doubt that more than a single star was situated within the CoRoT 
mask. Using a search for periodicity as a tool, our aim is to 
disentangle the composite 
light curve and identify the type of sources behind the variability.
}
 { 
Both flux and magnitude light curves were used. Conversion was applied
after a jump- and trend-filtering algorithm.
We applied different types of period-finding techniques including
MuFrAn and Period04.  
}
{
The amplitude and phase peculiarities obtained from the independent analysis
of CoRoT $r$,  $g$, and $b$ colours and ground-based follow-up photometric observations 
ruled out the possibility of either a background monoperiodic or a
Blazhko type RR\,Lyrae star being in the mask. The main target, an active star, 
shows at least
two spotted areas that reveal a $P_{\mathrm{rot}} = 8.8$~hours $(f_0 = 2.735$~c~d$^{-1})$ mean
rotation period. The evolution of the active regions helped to derive a  
period change of $dP/dt = 1.6\cdot 10^{-6}$ (18~s over the run) and 
a differential rotation of $\alpha = \Delta\Omega/\Omega = 0.0074$. 
The $0\fm 015$ linear decrease and a local $0\fm 005$ increase in the dominant period's amplitude 
are interpreted as a decay of the old spotted region and an appearance of a new 
one, respectively.  
A star that is detected only in the CoRoT $b$ domain shows a $f_1 = 7.172$~c~d$^{-1}$ 
pulsation connected to a $14\fd 83$ periodicity via an equidistant triplet
structure. The best explanation for our observation is a $\beta$ Cep star with a 
corotating dust disk. 
}
{}


\authorrunning{M. Papar\'o et al.}
\titlerunning{CoRoT\,102781750: periodicities in contaminated colour light curves}
 
\maketitle

%

\section{Introduction}

The French-led CoRoT space mission \citep{Baglin06,Auvergne09}, dedicated to 
asteroseismology and exoplanetary research, allows us to discover new 
details of the inner layers of pulsating stars, exoplanets, and many other exotic 
objects. CoRoT collects photometric time series data 
from adjacent stellar fields: it observes a few (defocused) bright stars ($<10$) 
in the seismofields and a large number of fainter stars ($\sim 10^4$) in the exofields. 
Data obtained both on the seismo- and exofields give us the possibility to broaden our knowledge 
on pulsating stars. The signal-to-noise ratio one can obtain for the bright seismofield 
targets (brightness range of 5.5--9 magnitude) allows detailed investigation of the 
low-amplitude, non-radial oscillations of the main sequence stars. 
Excellent papers on new results have been
published in the special CoRoT issue of A\&A on solar-like oscillation 
(HD\,49933 \citealt{Appourchaux09, Benomar09}, HD\,181906 
\citealt{Garcia09}), on $\delta$\,Scuti stars (HD\,50844 \citealt{Poretti09}), 
on $\beta$\,Cephei stars (HD~180642 \citealt{Degroote09}), and on Be stars 
(HD\,49330 \citealt{Huat09}; HD\,50209 \citealt{Diago09}). 

 \begin{figure}
 \centering
\includegraphics[angle=00,width=5cm]{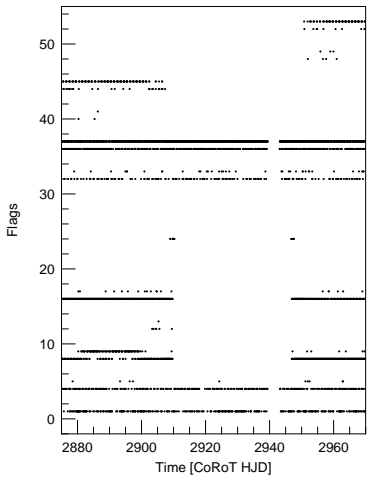}
 \caption{Distribution of non-zero flag measurements in the raw data of CoRoT\,102781750: 
energetic particles (1), crossing of SAA (4 and 32), Earth eclipse inward (8) 
or outward (16), and their combinations. After removal, 70.2\% of the original
measurements were analysed.
}\label{flag}
  \end{figure}

It was recognized early on that a number of research fields in stellar physics 
can benefit from the extremely precise time-series photometry of the exofield 
targets (magnitude range: 11--16) \citep{Baglin00, Weiss04}. The less frequent 
(8 min) default sampling combined with the large number of collected photons is ideal for
monitoring variable stars with periods longer than a few hours.
Unfortunately, up to now only a few dominantly radially pulsating RR\,Lyrae 
stars have been classified in the exofields. This is not surprising, however, since 
the CoRoT exoplanet fields are biased toward cool main sequence stars due to 
the CoRoT target selection procedure \citep{Debosscher09}.
The recently published CoRoT target, 
\object{V1127 Aql} \citep{Chadid10}, is the first RR\,Lyrae star continuously 
observed over 5 consecutive Blazhko cycles. A previously unknown complexity of 
the side lobe structure, up to the sepdecaplet structure, was identified within a total 
amount of 450 peaks in the frequency spectrum.  The CoRoT data of the Blazhko 
type RR\,Lyrae star, \object{CoRoT 101128793}, which has recently been investigated 
\citep{Poretti10}, suggest the presence of some non-radial modes besides the 
two radial modes and their linear combinations. 

The \object{CoRoT 102781750} 
was classified by the automatic CoRoT Variable 
Classifier (CVC, \citealt{Debosscher09}) to be an RR\,Lyrae star with a probability 
of 67\%, SPB type with 25\%, and Be star with 6\%. The uncertainty in classification
suggests contamination of neighbouring stars. Due to the high probability of the star 
belonging to the RR\,Lyrae class, the data of CoRoT\,102781750 was assigned to the 
CoRoT RR\,Lyrae working 
group \citep{Chadid09}. Given that all classifications for the star
represent pulsation, we used a periodicity search 
as a tool for disentangling the contaminated light curve of CoRoT\,102781750. In our 
analysis we benefitted from the brightness of the target ({\it B\,}=14.9, 
{\it V\,}=14.1), which allowed us to have `CoRoT colours' for the contaminated star. We attempted 
to disentangle the composite light curve and find the type of the sources behind the variability. 

\section{Observations and data processing}

To achieve the scientific goal of the exoplanet search programme, thousands of 
stars were observed simultaneously. The continuous light curves supplied 
a photometric precision that was typically a factor 100 better than Hipparcos and OGLE 
data \citep{Sarro09}. The CoRoT target, 
CoRoT\,102781750 ($\alpha = 6^{\mathrm h}45^{\mathrm m}29\fs 23$, 
$\delta = +0\degr 23\arcmin 4\farcs 99$) was 
observed during the first Long Run in the anti-centre direction (LRa01) from 
October 24, 2007 to March 3, 2008, i.e., over a time span of 131 days. 
According to Exo-Dat\footnote{http://lamwws.oamp.fr/exodat/}, the CoRoT/Exoplanet input catalogue 
\citep{Deleuil09}, its brightness and colour index are {\it B\,}=14.939 and 
{\it B}$-${\it V}=0.841, and its spectral type is K2V.
The Exo-Dat catalogue
gives $T_{\mathrm{eff}}$=5050~K for the temperature and {\it E(B$-$V)}=0.05
for the colour excess. More photometric data are given in Table~\ref{tab_1}. 

\subsection{Raw data}

 \begin{figure}
 \centering
\includegraphics[angle=00,width=7cm]{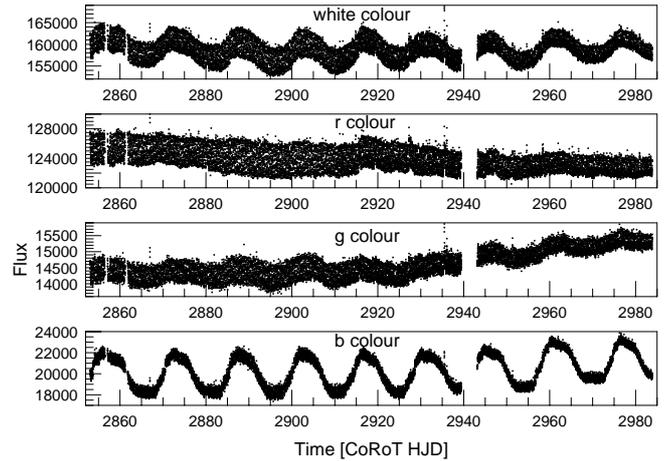}
 \caption{Zero-flag flux curves for the different colours. White flux was kept 
constant. The trends in the colour curves are the redistribution of flux over 
the colour borders due to the depointing fluctuations of the satellite.
}
 \label{flux}
 \end{figure}

With the 8-minute time sampling, generally used on the exofields, 
21\,219 measurements were gathered and delivered to us on N2 (science grade) 
level after reduction with the CoRoT pipeline  
\citep{Samadi06, Auvergne09}. The pipeline (in our case pipe\_version: 0.6) contains 
corrections for electronic offset, gain, electromagnetic interference, outliers, 
background, and jitter. Additionally, flags were attributed (N2\_version: 1.4) to 
measurements affected by one of a few instrumental effects (e.g., energetic 
particles (1), crossing SAA (South Atlantic Anomaly) (4 and 32), perturbation by 
Earth eclipse inward (8), or outward (16)). The distribution of the measurements
with non-zero flags is presented in Fig.~\ref{flag}. CoRoT time is given as HJD$-$2\,451\,544.5. 
Different flag values are the combination of these basic effects (e.g. 36=32+4 or 
37=32+4+1 at our case). It is quite understandable that high-energy particles and 
the disturbance because of the SAA can appear at anytime during the run. However, the 
effect of the Earth eclipse has no impact in the middle of the observing run over 
30--35 days. In our analysis we only used the valid zero-flag measurements (71.4\%).

\begin{table*}
\begin{flushleft}
\caption{ Photometric data of the CoRoT targets and possible contaminating stars (B-F). 
C102781750 and A are the same stars. $B$ and $V$ denote 
Johnson magnitudes, {\it r'\/} and {\it i'\/} stand for Sloan magnitudes, {\it JHK} are 2MASS near-infrared 
magnitudes, $B_1$, $R_1$, $B_2$, $R_2$, and $I$ are USNO B magnitudes obtained in different epochs, 
while the column denoted by `Sp.' contains the spectral classification. 
X and Y are the calculated target pixel coordinates on the CCD, while $\Delta x$ and $\Delta y$
are the pixel distances from the main targets. $\Delta$ is the angular distance in arcseconds 
measured from the main target, A. The superscripts b) and c) refer to the distance to stars B and C.}\label{tab_1}
\centering
\begin{tabular}{llllllllllllrrr}
\noalign{\smallskip}
\hline\hline
\noalign{\smallskip}
Star & {\it B} & {\it V} & {\it r'}& {\it i'} & & {\it J} & {\it H} & {\it K} 
& Sp. & {\it X} & {\it Y} & $\Delta x$ & $\Delta y$   & {\bf $\Delta$}
\\
 & mag & mag & mag & mag & mag & mag & mag & mag &  & pixel & pixel & pixel & pixel &  arcsec \\
\noalign{\smallskip}
\hline
\noalign{\smallskip}
C102781750 & 14.939 & 14.098 & 13.665 & 13.156 & & 12.314 & 11.811 & 11.692 & K2V   & 1165 & 1934 &  --   & --  & --\\ 
C102780084 & 16.066 & 15.038 & 14.542 &   --    & & 12.456 & 11.852 & 11.665 & G0III & 1178 & 1920 &  +13 & $-$14 & 44.3 \\ 
C102781578 & 18.426 & 17.211 & 16.576 &   --    & & 14.777 &  14.129 &  13.981 &   --   & 1185 & 1935 &  +20& +1 & 46.5\\
C102780066 & 17.622 & 16.839 & 16.608 &   --    & &   --    &   --    &   --    &   --   & 1182 & 1922 &  +17 & $-$12 & 48.3 \\
\noalign{\smallskip}
\hline
\noalign{\smallskip}
Star & {\it B$_1$} & {\it R$_1$} & {\it B$_2$} & {\it R$_2$} & {\it I} & {\it J} & {\it H} & {\it K} 
& Sp. & {\it X} & {\it Y} & $\Delta x$ & $\Delta y$  & $\Delta$\\
\noalign{\smallskip}
\hline
\noalign{\smallskip}
A    & 14.91 & 13.86 & 14.87 & 14.28 & 13.48 & 12.314 & 11.811 & 11.692 &  K2V  & 1165 & 1934 & -- & -- &  --\\
B+C  & 18.94 & 16.17 & 17.55 & 15.99 & 15.99 & 15.668 & 14.989 & 15.117 &  K4V  & 1167 & 1928 & +2 & $-$6 &  $8.4^{b}$\\
B+C  & --     &  --    &  --    &  --    &  --    & 16.158 & 15.338 & 15.248 & K4III &  --   &  --    &  --  &  -- & $11.6^{c}$ \\
D    & 18.67 & 18.36 & 18.27 & 17.98 & 17.98 &   --    &   --    &   --    &   --   & 1161 & 1930 & $-$4 & $-$4 &  13.1\\
E    & 19.54 & 14.87 & 18.55 & 17.81 & 17.33 & 16.355 & 15.878 & 15.637 & K5V?   & 1160 & 1936 & $-$5 & +2 &  12.4\\
F    & 19.74 & 17.35 & 18.96 & 17.37 & 16.98 & 15.721 & 14.998 & 14.608 & M3III & 1165 & 1938 & 0 & +4 &  9.3\\
\noalign{\smallskip}
\hline
\end{tabular}
\end{flushleft}
\end{table*}
After removing the obvious outliers,
about 14\,900 measurements (70.2\% of the raw data) were used in our analysis.
Usually for other CoRoT targets only about 10\% non-zero flag measurements
are obtained. We analysed the non-zero flag measurements to find any
regularity (e.g., binarity), but we failed. We cannot pinpoint any obvious reason
for discarding such a high percentage of the raw data for our analysis. 

\subsection{CoRoT colours}

In exoplanet research, the distinction between intrinsic stellar brightness variations (chromatic) 
and a planetary transit (achromatic) is a critical point. 
Therefore, for stars on the exofield chromatic information is taken with a bi-prism 
located close to the exoplanet CCD. Because of the prism and the astigmatism aberration, 
the point spread function (PSF) is a very low-resolution spectrum \citep{Rouan99}. The PSF 
exhibits an elongated shape with a typical size of 15$\times$10 pixels, 
which corresponds to about $35\times 23$ 
arcseconds. The optimal photometric mask used in the photometric on-board reduction (so-called 
template) fits the unusual shape of stars' images. To this end the assigned template consider
the temperature and colour indices of the target, its position on the CCD, and the 
position of possible contaminating stars. The template is selected from a 
set of 256 predefined templates optimizing the signal-to-noise ratio of the 
target's integrated flux \citep{Barge06}.

The two prisms, each made with a proper glass 
and angles, so that the mean wavelength is not deviated but the blue 
extends to the left and the red to the right \citep{Leger06}, enable spatial 
discrimination of contaminated stars \citep{Barge08}.
The spectrum is split by a proper selection of pixels into three spectral bands 
whose fluxes are recorded independently. 
The boundaries between the colours 
are arbitrary, and the final setting was based on stellar field 
simulations in the preparatory phase. To get the best spatial discrimination 
for contaminated planetary transit light curves, the 
following optimum distribution was set: 20\% of the bluest photons are assigned to 
the blue channel and 65\% of the reddest photons to the red channel. The remaining photons 
constitute the green flux, which is transmitted independently as well, in order to 
measure the total flux \citep{Rouan99}. However, the actual fraction of the colour fluxes 
depends on the target. The white light is kept constant over the observing run.
The coloured fluxes are very sensitive to depointing fluctuations of the satellite, 
not only because of the lower number of pixels but also because of the chromatic 
contamination \citep{Samadi06}. Any shift in the spectrum will transfer signal 
across the border from one colour to the other. Any change in the flux measured by 
each channel corresponds to a redistribution of the flux within the mask, since the 
total flux remains constant.

CoRoT\,102781750 was observed in colours. The 
photometric mask (templ\_ID:91) was 11 pixels long in the S-N ($X$) direction 
and 7 pixels wide in the E-W ($Y$) direction. The first position of the window
was 1165 and 1934 pixels in the $X$ and $Y$ directions, respectively.
The total number of pixels in the template is 63. Out of that, 37 (59\%) are 
attributed to the red, 7 (11\%) to the green, and 19 (30\%) pixels to the blue
region. The location of the right edge of the blue part is at four pixels, 
while the location of the left edge of the red part is at six pixels.

The zero-flag data are presented in Fig.~\ref{flux}. The white fluxes 
(top panel) do not show any trend. The decrease in red flux  
(2695 units) and the increases of the green ({\it g}) (1075 units) and blue 
({\it b}) colour fluxes (1273 units) can be attributed to the redistribution 
of fluxes over the colour border. By calculating moving averages, the 
redistribution was localized at around CoRoT~HJD 2924. The colour ratios to 
the total flux are 79\% in {\it r}, 9\% in {\it g}, and 12\% in {\it b} colour 
for our target. A similar distribution of flux (73\% in red, 11\% in green, and 
16\% in blue) in a K0V star into 3 channels is given by \citet{Leger09} in his 
Fig.~10.

The apparently different shapes of colour light curves cannot originate 
in the same (either pulsating, eclipsing, or active) star, therefore suggesting 
a severe contamination and needing different data processing for each colour.

Since we were primarily interested in the RR\,Lyrae type variation clearly
seen in the {\it r} colour, we removed the long-period variation from the {\it g} 
and {\it b} colours. Due to the different appearances of the {\it g} and {\it b} 
colour light curves, the data processing had to be done differently. In the {\it g} 
colour, the long-period variability only has a 0.02 mag amplitude compared to the 
dominant behaviour in the {\it b} colour which amounts to 0.2 mag. A moving boxcar 
method \citep{Chadid10} 
was used to get rid of the trends and jumps in the {\it r} colour. At the same step the 
fluxes were converted to magnitude. A similar step was used for the fluxes in the {\it g} 
colour using proper widths for the boxcar. However, this method did not work in the 
{\it b} colour. No proper value was found for the boxcar to get rid of the long-period 
variability. Finally we applied a preprocessing of the fluxes. Three frequencies 
(0.0674, 0.0563, and 0.0777 c~d$^{-1}$) were prewhitened from the fluxes in order 
to remove the large amplitude variation, then the trend- and jump-filtering algorithm and 
the magnitude conversion was used for the residual flux curve.

\section{Contamination of CoRoT\,102781750}\label{Contam}

\begin{figure}
 \centering
\includegraphics[angle=00,width=9cm]{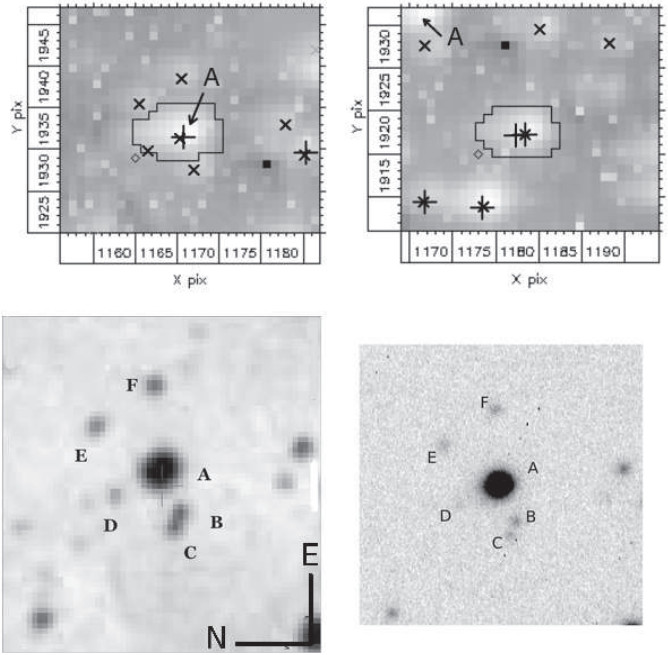}
 \caption{Map of the possibly contaminating stars CoRoT\,102781750 (upper left), 
CoRoT\,102780084 (upper right) as seen by the CoRoT satellite. 
(Position of stars contained in the Exo-Dat 
database and USNO-A2 catalogue are marked by plus signs `+' 
and crosses `x', respectively). 
The closer neighbourhoods of  CoRoT\,102781750 in
DSS (lower left) and Konkoly observation (lower right) are shown. The star
CoRoT\,102781750 is marked by A in each panel.
Data of the stars labelled by A-F are given in Table~\ref{tab_1}.
}
  \label{maps}
  \end{figure}

As a result of the CoRoT preparatory work, i.e., the target selection process, the 
Exo-Dat catalogue contains all the information on the target stars, as well as on their 
environments. This allows us to know the contamination of the stars in the fields in advance. 
For CoRoT\,102781750, a low contamination value of 0.0240795 is given in the catalogue. 
Three objects are mentioned in the vicinity of our target. For the sake of 
clarity their photometric parameters, positions on the CCD, and their relative 
distance to the target star are given in the upper part of Table~\ref{tab_1}. They are 
fainter than the target and are not closer than 40{\arcsec} (the CoRoT scale is  
$2 \farcs 32$/pixel, see the last column of
Table~\ref{tab_1}). The photometric masks in the two top panels of Fig.~\ref{maps} 
(left: CoRoT\,102781750, Win\_ID:~2539, templ\_ID:~91 and right: 
\object{CoRoT 102780084}, Win\_ID:~1615, templ\_ID:~138) clearly show that 
indeed the masks of the close-by Exo-Dat objects do
not overlap, and  no contamination is caused by these far-away stars.  
The low contamination level given by the Exo-Dat catalogue is
consistent with this fact. However, the different appearance of the colour 
light curves suggests severe contamination. 

The lower panels in Fig.~\ref{maps} show the closer vicinity of our target (A) on a DSS frame 
(left) and on a CCD frame obtained at the mountain station of Konkoly 
Observatory (right). The stars B, C, D, E and F are situated at or near the 
edge of the photometric mask, marked by ``x'' on the top left panel. Their 
photometric parameters and positions in pixels (obtained from the template) are 
given in the lower part of Table~\ref{tab_1}. Although the most nearby stars are faint, 
they are definitely candidates for possible sources of contamination,  
because, due to the prism, stars outside of the CoRoT mask can contribute to the
measured flux. 
The relatively large PSF of CoRoT with larger axes approximately in the S-N direction 
and the large aperture mask (in our case $11\arcsec \times 7\arcsec$) raise the possibility that 
some of the neighbouring stars may contaminate the light curve of CoRoT\,102781750.

Fortunately, a complete set of {\it J}, {\it H}, {\it K} values from the 2MASS 
catalogue \citep{Skrutskie06} was available, enabling a raw estimation 
of the spectral types of the neighbouring stars. Comparing {\it J$-$H}, {\it H$-$K} and {\it J$-$K} 
colour indices to the values of \citet{Bessel88}, our conclusion is that the stars marked with B, C and E 
are mid-K type main sequence or giant stars, while F has an even later spectral type, 
being an M3 giant. We did not find any {\it HJK} photometry for star D, since it is the 
only blue star in the vicinity of our target. The lower part of Table~\ref{tab_1} contains 
blue and red  photometry taken from the USNO\,B catalogue at two different epochs and 
in different passbands ($B_1$, $R_1$, $B_2$, $R_2$) (\citealt{Monet03,Sesar06}).  The accuracy of 
the catalogue is low, but not worse than $0\fm 5$. The star marked with E shows a rather 
unusual photometric behaviour, which may be an error.

As part of the ground-based follow-up programme, a limited number of control 
measurements ({\it BVRI}) were obtained with a Princeton Instruments VersArray:1300B 
back-illuminated CCD camera attached to the 1-m RCC telescope on January 
11, 2010 at Piszk\'estet\H o, the mountain station of Konkoly Observatory. 
A 200$\times$200 pixels window of the combined CCD frames (with a 0.3033 arcsec/pixel 
resolution) is given in the lower right panel of Fig.~\ref{maps}. Comparing the 
instrumental colour indices ({\it B$-$V}, {\it V$-$R}, {\it R$-$I}), we can confirm our 
estimation for the spectral types based on 2MASS colours. The stars marked 
with B, C, and F have colours similar to our target 
star (A). D is a very blue star, while the star marked with E has unusual 
colour indices: its {\it B$-$V} index suggests a very red star,
while the {\it V$-$R} and {\it R$-$I} indices do not confirm the
extremely red colour. There is no definite conclusion on the spectral type of star E.

To further constrain the nature of the possible contaminating stars, 
we obtained time-series photometric observations with the same camera and telescope, 
but without filter, on March 6 and 8, 2011 at 
Piszk\'estet\H o. Three hours of observations were gathered on both nights, which 
would allow us to detect RR Lyrae type variation in any of the targets despite the 
rather large seeing (3--5\arcsec). However, the stars A-F were constant within the error 
on these nights. The upper limit of the variation is 0.01 mag for the bright star A, 
and 0.05 -- 0.1 mag for the fainter targets (B-F). 

Due to their position to the left of the main target, there are two candidates
that may contaminate the CoRoT blue colour of the main target: a blue star (D)
and a red star (E). The photons from both stars (even the red part of their 
flux) can be measured in the {\it b} colour channel of the main target due to the 
prisms in the optical path. The closest candidate is definitely the star (D) with
spectral type B.

\section{Frequency analysis}

A standard frequency analysis was done independently for each colour with the 
software package MuFrAn \citep{Kollath90}. Investigations for amplitude and 
phase variability, and for significance and errors were obtained with 
Period04 \citep{Lenz05}. 

 \begin{figure}
 \centering
\includegraphics[angle=00,width=6cm]{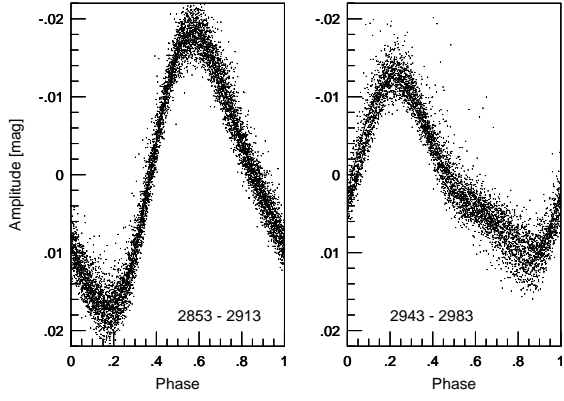}
 \caption{Light curve in the $r$ colour folded with the main pulsation period separately 
for the beginning (left) and the end (right) of the observing run. Continuous 
and dashed lines show the first and last cycles of the subset.
}
  \label{rdate}
  \end{figure}

\subsection{Results for the {\it r} colour}

\subsubsection{Fourier parameters}

The CoRoT {\it r} colour exhibits the RR\,Lyrae type pulsation in the clearest way.
However, the usually accepted multifrequency analysis, improving all the 
Fourier parameters at the same time in each step, turned out to be meaningless 
for the whole data set. After finding the dominant peak and its first harmonic, 
a third frequency appeared so close to the dominant frequency that the 
amplitudes were influenced by each other. The amplitude of the main frequency 
decreased compared to the single peak solution and the third frequency's 
amplitude was unrealistically high. This is a typical symptom of 
unresolved closely-spaced frequencies or of a real amplitude variation of a single 
frequency. In any case, the standard multi-frequency solution cannot be used to 
find the realistic Fourier parameters of more than two frequencies.

 \begin{figure}
 \centering
\includegraphics[angle=00,width=6cm]{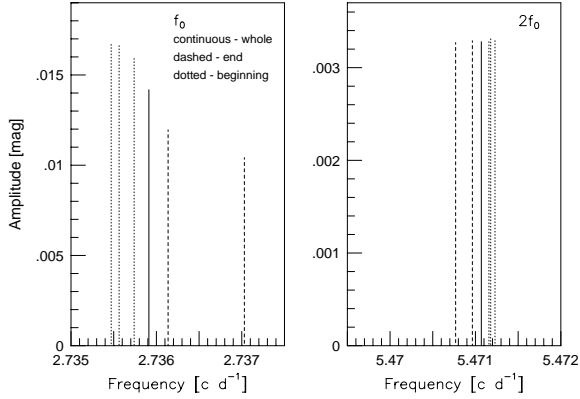}
 \caption{Frequency solution for five subsets and the whole data for the main 
frequency (left) and for its harmonic (right). The harmonic shows a constant amplitude, 
while the dominant mode's amplitude is obviously decreasing.
}
            \label{freqs}
  \end{figure}

The whole data set was divided into subsets in order to investigate a
possible amplitude variability. The separation of the subsets was connected to 
special features in the raw data as shown in Fig.~\ref{flux}. (CoRoT~HJD 2853--2939, 2853--2916,
2853--2913, 2943--2983, and 2916--2983). A jump appeared at CoRoT~HJD 2913 and a gap 
at CoRoT~HJD 2943. Through such a subdivision, any possible technical origin can be 
excluded or traced.

\begin{table*}
\begin{flushleft}
\caption{Results of the frequency analysis of CoRoT\,102781750 for each colour.  The parameters 
of $f_0$ and $2f_0$ are obtained from multifrequency analysis. Step-by-step 
frequencies are derived after prewhitening in each step.}\label{tab_2}
\centering
\begin{tabular}{llllllllll}
\noalign{\smallskip}
\hline\hline
\noalign{\smallskip}
CoRoT~HJD & 2853--2913  &  &  & 2853--2983 &  &  & 2943--2983 &  &    \\
Interval & Beginning    &  &  & Whole data  &  &  & End         &  &    \\
\noalign{\smallskip}
\hline
\noalign{\smallskip}
& Frequency & Amplitude & Phase & Frequency & Amplitude & Phase & Frequency & Amplitude & Phase  \\
&  c d$^{-1}$ & mag & degree & c d$^{-1}$ & mag & degree & c d$^{-1}$ & mag & degree \\
\noalign{\smallskip}
\hline
\noalign{\smallskip}
{\it r} colour     & 2.73547 & 0.01669 & 202.19  & 2.73591 & 0.01418 & 197.51 & 2.73703 & 0.01042 & 154.51  \\ 
$f_0$, $2f_0$      & 5.47116 & 0.00328 & 211.15  & 5.47107 & 0.00328 & 212.45 & 5.47077 & 0.00327 & 223.44  \\
\noalign{\smallskip}
\hline
\noalign{\smallskip}
step-by-step & 2.75590 & 0.00119 & 244.64  & 2.73001 & 0.00237 & 263.69 & 2.71866 & 0.00103 & 254.69  \\ 
             & 2.71274 & 0.00050 & 155.85  & 2.74206 & 0.00146 & 108.88 &    -     &     -   &   -     \\
             & 2.69314 & 0.00047 &  52.71  & 2.75399 & 0.00081 & 233.90 &    -     &     -   &   -     \\ 
             & 2.67940 & 0.00034 & 126.78  & 2.73699 & 0.00047 & 319.56 &    -     &     -   &   -     \\
             & 2.78023 & 0.00034 &  44.32  & 2.72469 & 0.00047 & 171.69 &    -     &     -   &   -     \\ 
             &     -    &    -    &   -     & 2.76073 & 0.00045 & 200.64 &    -     &     -   &   -     \\
             &     -    &    -    &   -     & 2.70529 & 0.00035 & 284.14 &    -     &     -   &   -     \\ 
             &     -    &    -    &   -     & 2.71571 & 0.00035 & 104.39 &    -     &     -   &   -     \\
             &     -    &    -    &   -     & 2.68824 & 0.00032 &  77.81 &    -     &     -   &   -     \\ 
             &     -    &    -    &   -     & 2.74833 & 0.00025 & 254.99 &    -     &     -   &   -     \\ 
             &     -    &    -    &   -     & 2.77136 & 0.00025 & 129.13 &    -     &     -   &   -     \\ 
\noalign{\smallskip}
\hline
\noalign{\smallskip}
\hline
\noalign{\smallskip}
{\it g} colour     & 2.73549 & 0.02328 & 203.11  & 2.73595 & 0.02004 & 198.34 & 2.73698 & 0.01519 & 158.32  \\
$f_0$, $2f_0$      & 5.47102 & 0.00419 & 212.55  & 5.47103 & 0.00427 & 212.73 & 5.47072 & 0.00443 & 225.07  \\ 
\noalign{\smallskip}
\hline
\noalign{\smallskip}
step-by-step & 2.75529 & 0.00181 & 254.50  & 2.72995 & 0.00309 & 266.97 & 2.71744 & 0.00125 & 244.31  \\
             & 2.71337 & 0.00074 & 143.33  & 2.74265 & 0.00189 &  92.33 &     -    &    -    &   -    \\ 
             & 2.69178 & 0.00070 &  59.56  & 2.75362 & 0.00108 & 238.57 &     -    &    -    &   -     \\
             & 2.67723 & 0.00059 & 151.15  & 2.76074 & 0.00069 & 194.73 &     -    &    -    &   -     \\ 
             &    -     &    -    &   -     & 2.73749 & 0.00067 & 294.60 &     -    &    -    &   -     \\
             &    -     &    -    &   -     & 2.72428 & 0.00063 & 189.25 &     -    &    -    &   -     \\ 
             &    -     &    -    &   -     & 2.71530 & 0.00053 & 238.57 &     -    &    -    &   -     \\
\noalign{\smallskip}
\hline
\noalign{\smallskip}
\hline
\noalign{\smallskip}
{\it b} colour     & 2.73549 & 0.02275 & 204.30  & 2.73598 & 0.01981 & 199.32 & 2.73683 & 0.01563 & 167.07  \\ 
$f_0$, $2f_0$      & 5.47106 & 0.00392 & 209.18  & 5.47116 & 0.00394 & 207.79 & 5.47051 & 0.00415 & 234.46  \\
\noalign{\smallskip}
\hline
\noalign{\smallskip}
step-by-step & 2.75601 & 0.00172 & 244.51  & 2.73010 & 0.00276 & 263.73 & 2.71957 & 0.00143 &  38.91  \\ 
             &    -     &    -    &   -     & 2.74239 & 0.00164 &  99.50 &     -    &    -    &   -     \\
             &    -     &    -    &   -     & 2.75427 & 0.00123 & 235.18 &     -    &    -    &   -     \\ 
\noalign{\smallskip}
\hline
\end{tabular}
\end{flushleft}
\end{table*}

Figure~\ref{rdate} shows the folded light curves before the jump (left panel) and after 
the gap (right panel). The first and the last cycles of the subsets are also 
shown in the panels depicting the evolution of 
the light curve. There is no large difference at the beginning but a substantial 
change can be seen at the end. The amplitude of the light variation is obviously 
higher at the beginning and the shape of the light curve is remarkably distorted 
at the end. The early appearance of the distortion as a bump evolves to a broad 
minimum. It does not appear at a constant position as in monoperiodic RR\,Lyrae stars \citep{Paparo09}.
{\it We definitely can exclude the case of a monoperiodic RR\,Lyrae variation 
without any further check.} If we suppose that the distortion is due to the bump, 
the migration (earlier bump phase to light maximum and later to the minimum) 
contradicts the bump migration behaviour generally observed in Blazhko RR\,Lyrae stars \citep{Guggenberger06}.

\begin{figure}
 \centering
\includegraphics[angle=00,width=6cm]{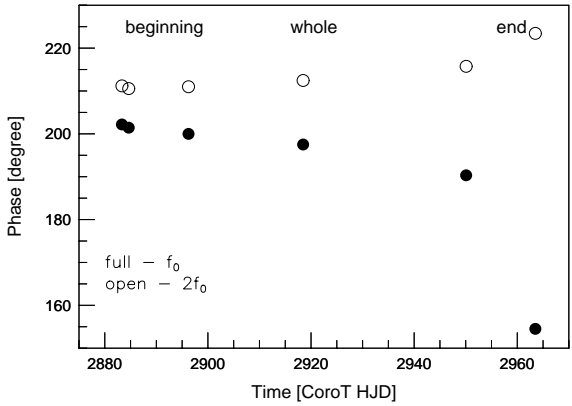}
 \caption{Phase behaviour of the main frequency and its harmonic for five subsets 
and for the whole run. The main frequency's phase reveals a sharp drop at the end of 
the run.}
  \label{phase}
  \end{figure}

The Fourier parameters of the main frequency and its harmonics are plotted for
the finally accepted subsets in Fig.~\ref{freqs} (frequencies and amplitudes) and in
Fig.~\ref{phase} (phases). The exact values are given in Table~\ref{tab_2} for the beginning
(left columns), for the end (right), and for the whole data set (middle). {\it The 
$P_0 = 0\fd 36551$ period fits the pulsation period of the Blazhko RR\,Lyrae 
stars. If the RR\,Lyrae type pulsation is attributed to a background star
contaminated by the main target, it is understandable why the amplitude of the 
pulsation is so low} ($0\fm 04$ in {\it r} and $0\fm 08$ both in {\it g} and {\it b}, the 
modulation amplitude is about $0\fm 015$) {\it and why we only see the first 
harmonic}.  A simple calculation shows that, when taking the USNO {\it B} and {\it R}
magnitudes into account (Table~\ref{tab_1}), the resulting low amplitude (and amplitude variation) is
consistent with the A+F or A+B+C blending scenarios.

Some additional effects can be obvious from the graphical representation
(Fig.~\ref{freqs}). As for the frequencies, the separate subset solutions are within 
the $\pm0.0038$~c~d$^{-1}$ error bar of the frequency resolution for the whole data set 
based on the Rayleigh frequency of a 131-day long observing run. Although the 
Rayleigh resolution is regarded in most cases as a dramatic overestimation of 
the real uncertainty (Kallinger et al. 2008), its application assures that our 
results are not over-interpreted. Nevertheless, the frequency values obtained at 
the beginning (dotted lines) or at the end (dashed lines) of the observing run 
show systematic arrangement around the whole data's solution (continuous line).
The higher frequency value at the end means a shorter period with an 18~s difference 
compared to the period at the beginning. At the same time the `let it free' 
solution 
of the first harmonic (when harmonics are determined as independent 
frequencies) 
shows much a narrower distribution with an opposite inner 
behaviour compared to the beginning and the end.

{\it If we regard the 18 s variation in the pulsation period as a period 
change,
a $dP/dt = 1.6\cdot 10^{-6}$ value is derived}. Apart from the abrupt changes, two 
types of continuous period change may be expected in RR\,Lyrae stars. Evolutionary 
period change is much slower \citep{LeBorgne07} than this value. The period 
change connected to the Blazhko modulation is usually faster than the value 
presented here, although slower modulations have been also reported \citep{Poretti01}.  

The amplitude variation of the main frequency and its harmonic is shown in Fig.~\ref{freqs}. 
While the amplitude of the main
frequency is steadily decreasing, the much lower amplitude of the harmonics
keeps a constant value. {\it No similar systematics were reported on 
Blazhko type RR\,Lyrae stars for the main period and its harmonic}.

The phases obtained with MuFrAn, using the frequency solution of a given
subset, are shown in Fig.~\ref{phase} for the main frequency (full circles) and for
its harmonic (open circles). The difference between the appropriate subset
values (with an epoch at CoRoT~HJD 2853 for each) is getting larger (from 10 to 70 degrees) 
close to 
the end of the run with a large drop in the main 
frequency's phase after the gap. Period change can be ruled out as an explanation, 
since only a period change of two orders of magnitude would cause the large phase change that is observed.

As control cases, the monoperiodic CoRoT RR Lyrae star (\object{CoRoT 101370131}) and the CoRoT Blazhko
type (V1127\,Aql) RR\,Lyrae star were investigated for their phase behaviour in subsets.
Neither the phase of the main period nor the first harmonic deviated more than
3--4 degrees in the monoperiodic star. In V1127\,Aql, the phase difference is
larger at maximum and only a few degrees at Blazhko minimum. The ever-changing
Blazhko type RR Lyrae CoRoT star 105288363 shows a periodic phase difference of
the main period and its first harmonic between only 29.8 -- 36.7 degrees connected
to the Blazhko period \citep{Guggenberger11}. No significant 
changes were found in the epoch-independent phase difference for RR\,Gem, 
a Blazhko type RR\,Lyrae star \citep{Jurcsik05b}. {\it The development of 
the bump in our target stems from the increasing phase difference between the main period and its 
harmonic}.  

To summarize, the amplitude ratio of the main 
frequency and its harmonic in the CoRoT {\it r} colour changes between 
$R_{21}$=0.196--0.313 from the beginning to the end with a $R_{21}=0.231$
value for the whole data set. The phase difference is changing between
9{\degr} and 68{\fdg}9 from the beginning to the end with a 14{\fdg}9 value for the 
whole data set. The equivalent epoch independent phases are  
$\phi_{21}$=4.482--6.360 in {\it r}, 4.473--6.256 in {\it g}, and 4.373--6.114 
in the {\it b} colours given in radian. 
For comparison, we present here the
$R_{21} = 0.417$, $\phi_{21} = 2.365$ values for Blazhko RR\,Lyrae stars obtained by the 
MACHO survey \citep{Alcock03}. 
A change of $R_{21}$ over the Blazhko cycle (between 0.4--0.55 for
\object{SS For} and 0.45--0.55 for \object{RR Lyr}) has been published 
\citep{Kolenberg09, Kolenberg06}.
{\it The amplitude ratio, the phase 
difference, and their variations with a possible Blazhko period do not concur with 
the general behaviour of the Blazhko RR\,Lyrae stars}.

\begin{figure}
 \centering
\includegraphics[angle=00,width=6cm]{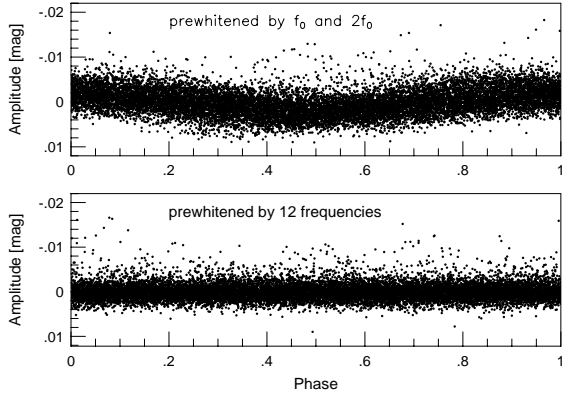}
 \caption{Folded residual light curves after prewhitening with 2 (upper panel) or
12 frequencies (bottom panel). The final residual scatter agrees
with the predicted value for a CoRoT target of similar brightness.
}
  \label{residual}
  \end{figure}

The residual light curve prewhitened by two frequencies still shows some 
variability as shown in the upper panel of Fig.~\ref{residual}. To get a residual with 
a constant average level, we continued a step-by-step frequency search
using the traditional significance criterium, $S/N > 4$ \citep{Breger93}. 
These frequencies are presented as separate sets in Table~\ref{tab_2}. 
The final residual light curve (lower panel in Fig.~\ref{residual}) has a zero average 
with a scatter $\pm0\fm 0019$, as predicted by \citet{Auvergne09} 
for a {\it V}=14$^{\mathrm m}$ brightness star ($0\fm 0015 - 0\fm 0016$).

Frequency separations reflecting a period comparable to the length of the data set 
and its fractions suggest that these peaks are only the manifestation of an 
overall amplitude variability of the main frequency during the observing run.
The number of step-by-step frequencies in a certain subset depends on the
complexity of the amplitude variation over the actual timebase. 

The first-order side peaks deserve more attention as they are rather asymmetric in 
amplitude. The {\it highest amplitude side peak appears on the right side} 
(period is shorter than the mean period by 233~s) {\it at the beginning, but 
at the end the left side peak} (period is longer than the main period by 213~s) 
{\it has the highest amplitude}.
Blazhko modulation leads to a complex side lobe structure from triplet      
(\object{AR Her}; \citealt{Smith99} or \object{RR Gem}; \citealt{Jurcsik05a}) 
to sepdecaplet structure (V1127\,Aql; \citealt{Chadid10}) in the Fourier spectrum. 
The side lobes are not exactly equidistant, and the left and right side peaks can 
differ in amplitude with the extreme case of a vanishing side lobe structure on 
one side \citep{Jurcsik05b}. 
Most Blazhko RR Lyrae stars investigated so far have a complex but static side 
lobe structure over a given Blazhko cycle.
Migration of the highest side peak from one side of the main frequency to the 
other on such a short time scale has never been
reported.

The $f_+/f_0$ ratio at the beginning is 1.007, 
$f_-/f_0$ is 0.9978 for the whole data and $f_-/f_0$ is 0.9933 at the end of 
the observing run.  
Frequency ratios between 0.95 and 1.05 were reported on Blazhko RR Lyrae
stars \citep{Moskalik03} in the OGLE database. Much closer frequency
ratios (between 0.99 and 1.01 and between 0.999 and 1.001 as well) were ascribed
to long and very long unresolved Blazhko effects \citep{Poretti01}.
However, a change in the
frequency ratio due to the migration of side lobes has never been reported.

The amplitude ratios are $A_+/A_0=0.0713$, $A_-/A_0=0.167$ 
and $A_-/A_0=0.0988$ for the beginning, the entire data set and the end, 
respectively. 
These values are at the lower limit of the 
$0.1 < A_{\pm}/A_0 < 0.3$ range obtained by MACHO data \citep{Alcock03}.

The $Q$ parameter \citep{Alcock03}, denoting the asymmetry in the triplet structure, is remarkably changing over the observing run (0.4 at the 
beginning and $-0.24$ for the whole data set but cannot be derived at the end
owing to the low value of $A_+$ reflecting an even higher negative value for 
the end than we obtained for the whole data set).
For MACHO RR Lyrae stars, the most probable distribution of {\it Q} lies in the 
$Q \approx 0.1$--0.6 range with a peak at 0.3 
\citep{Alcock00}.

{\it The spacing and the non-equidistance of the side lobes do not provide strong support for the possibility that 
one of our background stars is a Blazhko type
RR\,Lyrae star.  However, on the basis of these data, we cannot rule out the possibilty of a Blazhko star with a modulation period longer than the length of the observing run.} 
The modulation periods of Blazhko stars range from several tens to
several hundred times the pulsation period (ranges from 5.3--533~days)
(SS\,Cnc; \citealt{Jurcsik06} and \object{RS Boo}; \citealt{Nagy98}).

Although some of the representative parameter values ($f_{\pm}/f_0$, $A_{\pm}/A_0$ and {\it Q})
do not differ drastically from the typical value of Blazhko type RR Lyrae stars,
the other parameters ($R_{21}$, $\phi_{21}$, the side lobe values) and
especially their migration do not fit the background Blazhko type RR Lyrae 
scenario.

\subsubsection{Amplitude variation}

 \begin{figure}
 \centering
\includegraphics[angle=00,width=7.5cm]{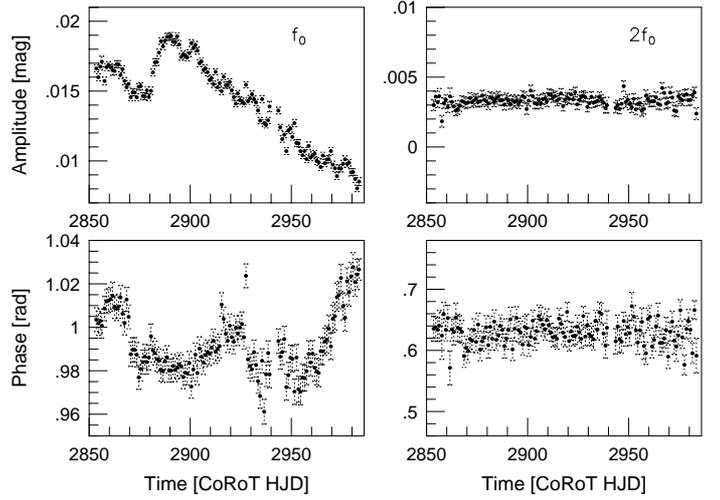}
 \caption{Fine-scale amplitude and phase behaviour of the main frequency (left) 
and its harmonic (right) obtained by Period04. The harmonic does not show amplitude 
and phase variation. A $0\fm 015$ amplitude decrease with a $0\fm 005$ local increase 
are shown in the main periodicity. The straight increase in the phase of $f_0$ is 
equivalent to the phase difference between the main frequency and its harmonic.
}
  \label{amp+phase}
  \end{figure}

In addition to the heuristic results obtained by the standard Fourier 
analysis of different subsets, the amplitude and phase variability on a 
fine scale (subdivided to one-day long time strings) was derived 
with Period04 and presented in Fig.~\ref{amp+phase}. Amplitudes (upper panels) and phases 
(lower panels) of $f_0$ and $2f_0$, along with their error bars, are plotted 
as a function of time. The harmonic does not show any variability either 
in amplitude or in phase. However, the main frequency exhibits a definite 
variation in amplitude. The total amplitude decrease is $0\fm 015$ over 
the whole observing run with a local $0\fm 005$ increase between CoRoT 
HJD 2880--2890. The dates of the local increase do not correspond to 
any of the mentioned technical effects. To be sure that the amplitude 
variability is not caused by any step in the data processing the flux 
values were also checked. The amplitude variability was also found with 
the same slope. 

An apparent amplitude variability can be caused by two closely spaced,
interacting frequencies \citep{Breger06}. In that case a similar
amplitude and phase variation appears with a certain phase shift between
them (it is 90{\degr} in the case of equal amplitudes).

Although the main frequency shows a structural phase variation over the
run (lower left panel of Fig.~\ref{amp+phase}), its appearance does not follow the amplitude
variability curve ruling out the closely spaced frequencies as a reason
of the amplitude variability. A finer structure of the phase behaviour is given 
because of the higher time resolution than in Fig.~\ref{phase}. The steady change at the 
end of the data set, from CoRoT~HJD 2950 onwards, agrees in value with the phase change shown in Fig.~\ref{phase} 
(taking into account the different definition of phase in MuFrAn and Period04).

The most remarkable result of this investigation is the steady decline of the
amplitude with a local increase. The slope of the amplitude variability is 
similar before and after the local increase. If we take the value of the amplitude
variability as a Blazhko modulation amplitude (half or one third of the 
pulsation amplitude), the prediction (large modulation amplitude for short 
pulsation period), is not broken. Based on a survey of about 900 RR\,Lyrae stars 
\citep{Jurcsik05b} $MAX(f_{\mathrm m})= 0.125f_0-0.142$ gives the shortest Blazhko 
modulation cycles, in our case it is 5 days. A longer Blazhko period than the 
length of the run cannot be excluded. {\it However, a 0.005 magnitude local 
increase in the Blazhko modulation amplitude has never been reported in Blazhko 
type RR\,Lyrae stars}.

\subsection{RR\,Lyrae type variation in {\it g} and {\it b} colours}

The RR\,Lyrae type variation shows similar behaviour in {\it g} and {\it b} to what we discussed
for the {\it r} colour. The result of the Fourier analysis is given in Table~\ref{tab_2}.
Both the frequency and amplitude ratios of the main peak and the first-order
side peaks are slightly different in value but similar 
in trend for {\it g} and {\it b} to what we discussed in the {\it r} colour. 
The amplitude variability is similar in all three colours 
(0.015, 0.017, and 0.017 values are in {\it r}, {\it g}, and {\it b}, 
respectively).
The main period's change is 17.2~s in {\it g} and 15.4~s in {\it b} towards shorter periods.
The first order side peaks are shorter by 227~s and 235~s at the beginning,
however, they are longer by 227~s and 200~s at the end in the {\it g} and {\it b} colours, 
respectively. The change in the mean period decreases from the {\it r} to 
{\it b} colour, reflecting the colour dependence,
but the first order side peaks behave in a similar way.

It is worthwhile mentioning the colour amplitude ratio of the main periodicity.
$A_g/A_r = 1.395$, 1.413, and 1.457, and $A_b/A_g = 0.977$, 0.989 and 1.029
values were obtained for the beginning, the whole data, and for the end of the 
run, respectively. The amplitudes are remarkably larger in {\it g} than in {\it r}
and practically the same in the {\it g} and {\it b} colours. Although the values do not
differ very much, there seems to be an increase from the beginning to the end
both in $A_g/A_r$ and $A_b/A_g$. 

Unfortunately, the CoRoT colours do not match the 
Johnson-Cousins {\it B}, {\it V}, {\it R}
bands exactly and they are not calibrated, so we cannot compare the $A_g/A_r$  
and $A_b/A_g$ ratios to the $A_V/A_R$ (=1.23) and 
$A_B/A_V$ (=1.30) values obtained for RR\,Lyrae stars \citep{Jurcsik06}.

\subsection{Additional periodicity in the {\it g} and {\it b} colours}

An additional $f_1 = 7.173$~c~d$^{-1}$ ($P = 3.3$~hours) periodicity, 
which is presented 
in Table~\ref{tab_3}, appeared in both the {\it g} and {\it b} colours. Its amplitude is close 
to the significance level in {\it g}.

\begin{table}
\begin{flushleft}
\caption{Additional frequencies in the {\it g} and {\it b} colours. A triplet structure spacing 
with the long-term periodicity is found. 
The rotational splitting of a non-radial mode is the most plausible explanation.}\label{tab_3}
\centering
\begin{tabular}{llll}
\noalign{\smallskip}
\hline\hline
\noalign{\smallskip}
& Frequency & Amplitude & Phase \\
& c~d$^{-1}$ & mag & degree \\
\noalign{\smallskip}
\hline
\noalign{\smallskip}
{\it g} colour     & 7.173299 & 0.00045 & 152.07  \\ 
\noalign{\smallskip}
\hline
\noalign{\smallskip}
{\it b} colour     & 7.17229 & 0.00468 & 177.90  \\ 
             & 7.10494 & 0.00196 & 172.03  \\
             & 7.23989 & 0.00097 & 290.85  \\ 
             & 7.18169 & 0.00096 & 300.52  \\ 
             & 7.16077 & 0.00063 & 307.64  \\
\noalign{\smallskip}
\hline
\end{tabular}
\end{flushleft}
\end{table}
 \begin{figure}[h]
 \centering
\includegraphics[angle=00,width=7cm]{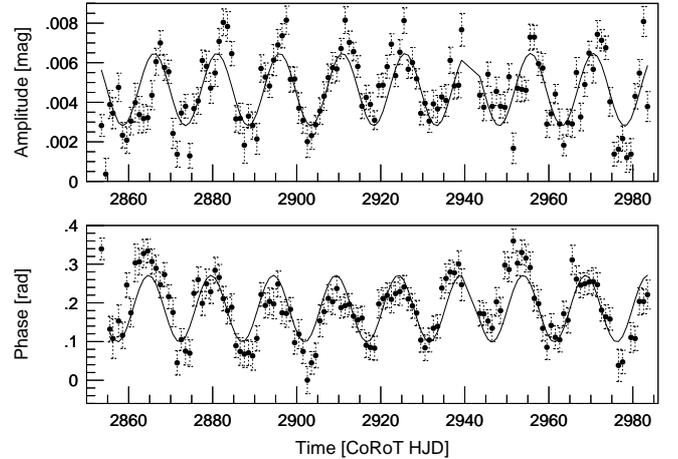}
 \caption{Long-term modulation in the amplitude and phase of $f_1$.  
Both time scales belong to the same star/stellar system.
}
  \label{f1}
  \end{figure}

However, it appears as the second largest amplitude periodicity with
a triplet structure in {\it b} (Table~\ref{tab_3}). The triplet is equidistant with a
0.067~c~d$^{-1}$ spacing reminding us of one of the frequencies that we used in
prewhitening of the flux curve in {\it b} colour (0.0674, 0.0563, and 0.0777~c~d$^{-1}$).
The two additional frequencies in Table~\ref{tab_3}, found above the significance
level with a +0.0094 and $-0.0115$~c~d$^{-1}$ spacing to $f_1$, are connected via a similar 
spacing to the other frequencies of the preprocessing step. However, it is
worthwhile emphasizing that the triplet structure of the additional periodicity
appearing in the light curve developed from the residual {\it b} flux curve. A
straightforward conclusion is that the triplet structure of the short-term 
pulsation and the long-term light variation, dominant in the {\it b} colour,
are connected.

The amplitude and phase variation of $f_1$ obtained with Period04 using a 
1-day long time resolution are shown with the error bars in Fig.~\ref{f1}. The phase 
shift between the maximum of the amplitude and phase ranges from 7\% to 13\% of the 
long-term periodicity. It is highly unlikely that the modulation
with the same period as the long-term variability is coincidentally caused by closely spaced
frequencies. A distinct period search around the maximum and minimum 
of the long-term periodicity confirmed that only the additional periodicity has 
a different amplitude, while the RR\,Lyrae type pulsation does not. The parameters of the 
single sine fits are given in Table~\ref{tab_4}. The imperfect fit (a need for the second 
periodicity) is clearly seen, especially in the phase curve (given in Table~\ref{tab_4},
too).

The close connection of the two types of variability strongly suggest that
they are the intrinsic light variation in a single star or a stellar system.

\begin{table}
\begin{flushleft}
\caption{Fitting parameters of the amplitude and phase modulation curves, presenting a new 
detailed approach to describing the triplet structure.}\label{tab_4}
\centering
\begin{tabular}{llll}
\noalign{\smallskip}
\hline\hline
\noalign{\smallskip}
& Frequency & Amplitude & Phase \\
& c~d$^{-1}$ & mag & degree \\
\noalign{\smallskip}
\hline
\noalign{\smallskip}
amplitude modulation     & 0.06698 & 0.00183 & 134.75  \\ 
\noalign{\smallskip}
\hline
\noalign{\smallskip}
phase modulation    & 0.06701  & 0.07872 & 219.79  \\ 
          & 0.05686 & 0.03327 & 274.36  \\ 

\hline
\end{tabular}
\end{flushleft}
\end{table}

\subsubsection{Long-period light variation}

Knowing all the `disturbing pulsations' in the {\it b} colour, we separated the
long-period light variation to derive its particular features. The original {\it b} 
colour flux data converted to magnitude were prewhitened with the frequencies 
given in Table~\ref{tab_2} for {\it b} colours (5 frequencies) and the triplet 
in Table~\ref{tab_3}. The resulting, clean long-term variation was analysed. 
The frequency analysis resulted in the same frequencies (see Table~\ref{tab_5}) 
as in the preprocessing phase, which is not surprising. Besides the frequency and period 
values, the amplitudes and phases are also given in Table~\ref{tab_5}. The periodicities 
have practically a 2- or 3-day equidistant spacing, taking an uncertainty in 
resolving three frequencies from a comparatively short timebase into account (nine cycles during 131 days). 
However, for a continuous data set, it is surprising that the spacing is exactly an integer 
multiple of the day. Similar uncertainties (cycle/day aliases) usually appear in ground-based,
single-site observations. We do not see any obvious technical reason for these aliases.

\begin{figure}[h]
 \centering
\includegraphics[angle=00,width=8cm]{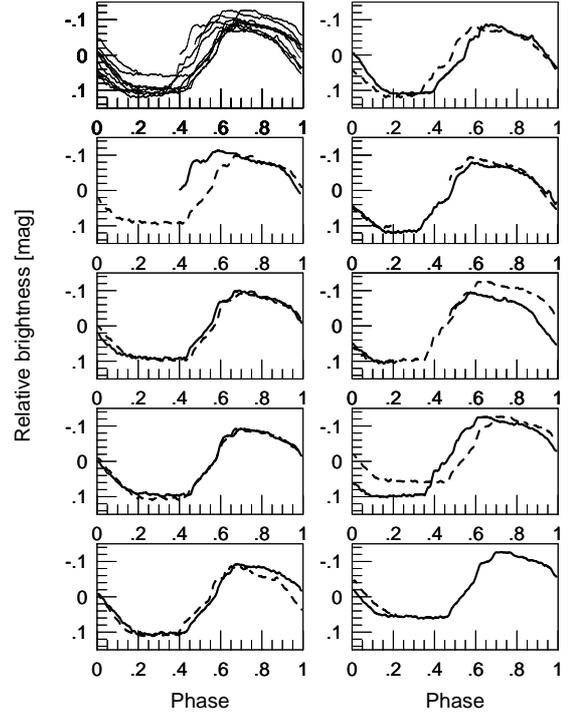}
 \caption{Long-term variability in {\it b} colour without the  pulsation. Consecutive 
cycles are not repetitive. The most remarkable feature is the unevenly long, flat part 
of the cycles. }
  \label{long}
  \end{figure}

After computing the moving average of 25 consecutive photometric points we folded the light 
curve with the 14.83 days period (Fig.~\ref{long}). The top left panel shows 
nine cycles together, demonstrating clearly that the cycles are not exactly 
repetitive. The lower left panels, continued on the right from top to bottom, give 
a detailed comparison of the consecutive cycles with overlaps. A given cycle is shown 
in two consecutive panels, first by a continuous and then by a dashed line.
Besides the small irregularities, some clear features can be easily identified.

\begin{table}[]
\begin{flushleft}
\caption{Fourier parameters of the long-period light variation. Due to the short 
timebase (9 cycles) the third value is obtained in a step-by-step analysis. The 
rotational period (Table 4) and the long term periodicity have the same values.}\label{tab_5}
\centering
\begin{tabular}{lllll}
\noalign{\smallskip}
\hline\hline
\noalign{\smallskip}
& Frequency & Period & Amplitude & Phase \\
& c~d$^{-1}$ & days & mag & degree \\
\noalign{\smallskip}
\hline
\noalign{\smallskip}
     & 0.06744 & 14.83 & 0.09571 & 143.78  \\ 
     & 0.05637 & 17.74 & 0.02136 & 217.14  \\ 
     & 0.07771 & 12.87 & 0.02033 & 272.96  \\ 
\noalign{\smallskip}
\hline
\end{tabular}
\end{flushleft}
\end{table}

The most remarkable feature is the practically constant part of the cycles at
minimum over the 0.2--0.3 phase range (with an increase after a decrease in 
length). The starting point of the  rising branch is rather sharp in contrast 
to the end of the descending branch. The cycles have a broad maximum with slightly 
decreasing brightness over a time interval similar to the constant part in the minimum. 
A similar regularity seems to exist in the decrease in the brightness and in the 
increase in its duration. It is worthwhile mentioning the horn or standstill 
appearing before the maximum brightness.

Finally, we conclude that the long-period variability is not strictly
repetitive from cycle to cycle, as we would expect in a simple eclipsing
binary system in which the variation is caused by a simple geometric effect.

\section{Discussion}

\subsection{RR\,Lyrae type pulsation: pro and contra}

We summarize our conclusions in favour of a background RR\,Lyrae star and 
against it.
The period we detected in the data agrees with the shortest period RRab stars. Its low amplitude
can be explained by the contamination. The lower amplitude and distortion 
of the light curve at the end of the run recalls Blazhko RR\,Lyrae 
stars. The Blazhko modulation period may be much longer than the length of 
the data set. The $f_{\pm}/f_0$ and $A{\pm}/A_0$ ratios are not exceptional. 

However, the local maximum seen on the descending branch of the light curve towards 
the end of the observing run (Fig.~\ref{rdate}) does not look like the bump in  
RR\,Lyrae light curves. The migration of the bump from an earlier 
pulsation phase (near maximum Blazhko phase 
-- highest light curve amplitude) to a 
later pulsation phase (near minimum Blazhko phase -- lowest light curve 
amplitude) does not follow the RR\,Lyrae scheme. The main periodicity 
varies (18--15~s) on a short timescale, presenting much faster ($dP/dt=1.6\cdot 10^{-6}$) 
period variation than evolutionary, but much slower than most Blazhko modulation changes. 
A large abrupt increase in the phase difference between the main frequency and its harmonics is not known 
in RR\,Lyrae stars. The highly constant amplitude of the first harmonic contradicts
the typical decrease found in Blazhko type RR Lyrae stars. The $R_{21}$ is much 
lower and $\phi_{21}$ is very different from the values obtained for either monoperiodic or for Blazhko 
RR\,Lyrae stars. The change in sign of the $Q$ parameter from the beginning 
to the end of the run; i.e., the different location of the higher amplitude 
side peak (from right (beginning) to left (end)) is a basic argument against 
the background RR\,Lyrae hypothesis. A local increase of 0.005 magnitude in 
the amplitude variation of the main frequency cannot be interpreted in the 
frame of the contaminated background RR\,Lyrae star.  
However, the low-amplitude light variation is 
a well-established fact.

Our strongest argument against the RR\,Lyrae interpretation is given by our
ground-based photometric observations taken on two nights (see Sec. 3.). These data essentially 
rule out RR\,Lyrae type variations in the nearby, possibly contaminating, stellar candidates
(B-F), as well as in any unresolved contaminating star in the immediate vicinity of our
main target (A).

\subsubsection{Stellar activity -- spotted star}

If the RR\,Lyrae type pulsation is not caused by a contaminated background 
RR\,Lyrae star, then we are faced with a more probable situation where the main
target, of spectral type K2, shows a light variation with a periodicity
similar to the period of an RR\,Lyrae pulsator. Late type stars with 
convective envelopes are all affected by magnetic processes that give rise 
to a rich variety of phenomena on their surface \citep{Donati06}. 
Thousands of stars were discovered showing rotational modulation due to 
star spots through different surveys (OGLE -- \citealt{Guinan97, Wozniak02}; 
ASAS -- \citealt{Pojmanski02}; MOST -- \citealt{Croll06, Walker03, Walker07}; 
CoRoT -- \citealt{Lanza09a, Lanza09b, Lanza10a, Lanza10b}).

If the low-amplitude variability is caused by surface phenomena, such as 
spots, then it reflects the rotational period of the star and the peculiarities
of the star spot region.  A variation of a few hundredths of a magnitude due to spots is 
typical for active stars. According to its light variation, CoRoT\,102781750 is more 
active than the Sun, like the CoRoT planet-hosting star 2a \citep{Lanza09a}, 
with $\approx$20 times greater modulation than that of the Sun during the 11 year 
cycle. Its activity is between CoRoT-Exo-2a (0.06) and CoRoT-Exo-6a (0.027) 
\citep{Lanza09a, Lanza10b}. The $P=0\fd 3655$ period is not exceptional if we regard 
it as a rotational period. The excellent review of star spots by \citet{Strassmeier09}
 presents a list of late type stars with Doppler images. Similar or even 
shorter rotational periods are listed for W\,UMa system (\object{YY Eri}), for single star
(RXJ1508.6) and for a K3V spectral type weak-lined T\,Tauri (WTTS) star 
(\object{HD 197890}: $T_{\mathrm {eff}} = 5000$~K, $P_{\mathrm {rot}} = 0\fd 380$, 
$v_{\mathrm {equ}} = 156.3$~km~s$^{-1}$, $R = 1.2 R_{\sun}$).

In this interpretation, the main period ($ P_0 = 0\fd 36551$) gives the rotational 
period of the average spotted area. The harmonic ($P_{\mathrm{rot}}/2$) may represent a 
second, equally-spaced spot group or the fact that the rotation of the averaged 
spotted area departs from a pure sine wave \citep{Queloz09}. The phase change 
between the main period and its harmonic ($\approx$60{\degr}) can be interpreted as a 
change in the separation of the two active longitudes.

The amplitude decrease of the light curve from the beginning to the end (Fig.~\ref{rdate})
can be attributed to a decay in a spotted region located at phase 0.8 (right
panel). Slightly before and definitely by CoRoT~HJD = 2943 a new spotted region 
appears at phase 0.5. The slight decrease in length of the main period 
($dP/dt = 1.6\cdot 10^{-6}$), caused by the evolution of the spotted region to a faster-rotating 
position (latitude) on the star, is comparable to the $dP/dt = 1.5\cdot 10^{-6}$ obtained for 
\object{UZ Lib}, interpreted as a probable indication of dependency on the activity 
cycle \citep{Olah03}.

Measurements of the rotational period of stars with spots allow the detection 
of differential surface rotations. If there are star spots at different latitudes, 
then the difference of the longest and shortest periods observed can be taken 
as an estimate of the averaged amount of surface differential rotation. Any kind of
one-dimensional, disk-integrated data, like time series photometry, cannot 
resolve the ambiguity of whether the polar or the equatorial regions rotate 
faster or slower \citep{Strassmeier09}.

Of course, the final conclusion has to be confirmed by other methods, such as Doppler
imaging or line profile analysis.  We can, however, come to some conclusion based on our 
own Fourier analysis and the distinct peculiarity of the continuous, high-precision 
CoRoT light curve producing a direct tracking of the active regions.

The Fourier analysis revealed the dominant period ($P_0 = 0\fd 365568$ and 
$P_0 = 0\fd 365359$ for the beginning and the end, respectively) of the highest 
concentration of the spotted area.  We found additional side peaks, which we can
attribute to the rotation of a larger independent spot. If we interpret the 
shorter (beginning: $-234$~s) and longer (end: +213~s) periodicity in the frequency 
spectrum as a sign of differential rotation, then by using $P_{\beta}/P_{\mathrm {eq}}=1-{\alpha}
\sin^2 {\beta}$ (see e.g. \citealt{Olah03}), where $P_{\mathrm {eq}}$ and $P_{\beta}$  are the 
rotational periods at the equator and at a latitude $\beta$, we get a minimal 
$\alpha = \Delta\Omega/\Omega = 0.0074$ differential rotation parameter by assuming 
$\beta = 90$\degr. This is less than one quarter of the differential rotation 
measured on the Sun (0.04--0.05 as given by \citealt{Lanza09b}). A similar value 
(0.0071) was found for the exoplanet hosting star, \object{CoRoT-Exo-2}a \citep{Lanza09a}. 
\object{CoRoT-Exo-6}a revealed a much higher ($0.12\pm0.02$) value \citep{Lanza10b}. An 
anti-solar differential rotation ($\alpha = -0.0026$) was determined for UZ\,Lib 
for the first time from the light curve alone by \citet{Olah03}, a result 
supported by Doppler images. Model calculations for \object{$\tau$ Boo} show that the 
equator-pole differential rotation is 0.15--0.18 \citep{Catala07}. The low 
differential rotation of CoRoT\,102781750 suggests that the spots are concentrated in a 
narrow belt. The right side peak (shorter period) at the beginning of the run can 
be attributed to a feature (spot) located closer to the faster rotating region of 
the star, either the equator in a solar-type case or the pole in an anti-solar type 
activity. The left side peak (longer period) at the end of the run is attributed to 
a new activity region appearing in the slower rotating region compared to the
highest concentration spotted area. A prograde and retrograde migration of the 
individual spots in the old and new active regions, respectively, or a differently 
oriented surface flow pattern would result a similar right-left side lobe structure.
The local increase in the amplitude between CoRoT~HJD 2880--2890 can be attributed 
to a larger area covered by star spots owing to the appearance of a new spotted region.

Since more observational facts fit the active star hypothesis, we conclude that 
CoRoT\,102781750 is not a background RR\,Lyrae star but a rotationally modulated, active star.
Our new ground-based photometry, with lower light variation than 0.01 magnitude,
excludes RR Lyrae type variability because a stable RR Lyrae pulsation, even with a Blazhko
modulation cannot disappear. However, it supports the active star hypothesis, as a variability 
caused by activity can decrease
or disappear in the course of the evolution of the active regions.

\subsection{Short-period pulsation}

In the section devoted to the contamination issue (Sec~\ref{Contam}.), 
we concluded that two candidates,
an uncertainly red (E, K5V?) and a definitely blue star (D) are responsible for 
the light variation in the CoRoT blue channel. Taking the 7.172~c~d$^{-1}$ 
frequency into account, we only have two options. 
The short-period pulsation belongs to a high-amplitude Delta 
Scuti star (HADS) or to a $\beta$\,Cephei type pulsation. We
do not know any late-type variable star that produces such a short periodicity 
(rotational modulation is also excluded). Since the late spectral type fits 
neither the HADS nor the $\beta$ Cephei stars, there is a higher 
probability that the short-period pulsation is shown by the blue star. 

The weak appearance of the first harmonic of the short-period pulsation at 
14.34503~c~d$^{-1}$ with a $\phi_{21} = 3.42 \pm 0.3$ rad phase value is in excellent
agreement with what is expected for HADS \citep{Poretti01}. 
However, the Konkoly colour photometry suggests a very
early spectral type compared to the mid spectral classes
of the HADS type. In addition, the triplet can not be explained
by rotational splitting in an HADS (radial mode).
Therefore equally spaced, non-radial modes have to be supposed.
The rotationally split non-radial 
mode in a $\beta$ Cep type star is the most plausible explanation for the 
short-period pulsation.
 
\subsubsection{Long-period variability}

The long-period variability has the largest amplitude ($0\fm 2$) amongst the
different kinds of variability presented here. Since in our interpretation 
the long-term variability is connected to a background star, it is worthwhile
checking the amplitude of the intrinsic variability. The {\it b} channel is
not contaminated by the whole flux of the main target, only by a fraction
of the {\it b} channel total flux. According to the flux variation in Fig.~\ref{flux},
roughly 4--5000 flux units can be attributed to the background star in the {\it b}
channel (roughly 1000 in {\it g} and there is no sign of the flux variation in {\it r}). 
Calculating the intrinsic amplitude dependence on the flux of the
background star, we are faced with a 0.7 magnitude intrinsic variation.
Usually a geometrical effect, the binarity can produce high variability. It 
is highly unlikely that the long-term variability belongs to the red star, 
since the short- and long-term variabilities seem to be connected, as the 
amplitude and phase modulation show. Both the short- and long-period 
variabilities belong to the same system or star.

According to the logical arguments there are two scenarios that may explain the observed variation.
In the first one, we are faced with a binary system containing a $\beta$ Cep type pulsating star. 
The system has to be synchronized since the rotational period (triplet spacing) 
and the orbital period are the same. However, the non-repetitive cycles and the 
long, constant minimum of the long-period variation do not support a binary system
with a simple geometry.

In the second scenario, the short- and long-period variations are produced in
a single star. The only possibility for a blue star is that we are faced with a
$\beta$ Cep/SPB hybrid pulsator in which a p mode (7.172~c~d$^{-1}$) and some g modes
(0.067, 0.0777 and 0.0563~c~d$^{-1}$) are excited. The best example of a hybrid star detected
before CoRoT is \object{$\gamma$ Peg} \citep{Chapellier06}. The p mode (6.01~c~d$^{-1}$)
fits our case, but the g modes (0.68 and 0.87~c~d$^{-1}$) represent a much shorter 
periodicity than our long-period variability. The excitation of modes with a
(high) radial order that fits the observed period is very unlikely. It would be a 
very improbable coincidence if an SPB star has exactly the same period as the
rotational period of a $\beta$  Cep star. Moreover, the light 
curve of the long-period variability does not look like an SPB light curve.

The periodic Be stars ($\lambda$\,Eri variables) show strictly periodic light
variations, but with periods in the range of 0.5--2.0 days. The shortest period 
is found in \object{HR 3858} (B6Ve, 6.7 hours), twice as long as our detected periodicity 
\citep{Balona95}. The review paper by \citet{Porter03} confirms the period
range. These stars appear to have very active photospheres and outbursts in which the 
brightness can increase or decrease by several tenths of a magnitude. The 
regularity of the long-period variability with such a long constant minimum and 
the modulation of the short-period variation can hardly be explained by outbursts.

Thus, no definite conclusion can be made with the information at hand.
However, we know the rotational period of the blue star given directly by the
rotational splitting, e.g. 14.83 days. The luminosity variation with a 14.83-day
periodicity must be something that rotates with the star. We have excluded the
synchronized companion, but a hybrid solution can be constructed. A non-radially pulsating 
$\beta$\,Cep star (rotational splitting is possible) surrounded by a corotating 
dust disk can explain the similarity of the modulation of the short period and
the orbital period. Higher density clumps in the disk orbiting at different 
distances from the star explain the periodicity at 14.83, 12.87, and 17.74 days. 
Maybe the instability of the modulation can also be explained by the 
non-homogeneous density distribution of the corotating disk. The disk is not
in the equatorial plane, but it has some inclination. The pulsating star is 
covered by the dust disk for a different time period from cycle to cycle. 
The life time $ < 10$~Myr of a young $\beta$\,Cep star does not exclude the 
existence of a dust disk around it, like in the Herbig Ae/Be stars. A fading 
by the circumstellar material with 0.7 magnitude is not unusual. However, a 
circumstellar disk would produce an infrared excess that we did not find in 
the 2MASS catalogue.

If the star E, which has an uncertain red spectral type and an unusual appearance
in its colours, is really an early-type star ($\beta$ Pictoris or UXOR) surrounded by
circumstellar material, this would provide an explanation for the
red colour.  However, the short-term periodicity would have to be explained in a
satisfying way.

Our ground-based observation did not show any variation in the stars D and E.
However, the low amplitude, short-period variation has not been expected to be
observable in such a faint star. The light variation of the long term variabilty
has a 3 to 4 day long part that is constant at minimum and a slightly decreasing part around
maximum.  Two nights of observation over three nights did not contradict our
hypothesis.

A separate and detailed observation of the candidates, ideally with spectroscopy 
(a great challenge due to the targets' faintness), will help to solve this puzzle.

\section{Conclusions}

We conclude that CoRoT\,102781750 is a spotted star with an 8.8-hour average
rotational period.  A $dP/dt = 1.6\cdot 10^{-6}$ period decrease was derived from
 the evolution of the spotted region. A differential rotation of $\alpha = 0.0074$
is measured. An alternate (shorter or longer) second periodicity was 
interpreted as an independent spot or a new active region at the faster
or slower rotating part of the star, respectively. The $0\fm 015$ decrease
in the light variation is due to the decay of the spots, while the local
$0\fm 005$ increase is a sign of a higher covered area due to the appearance 
of a new spotted region.

The CoRoT mask contains flux coming from a background star. Short-term
variability (7.172~c~d$^{-1}$) and a long-term variability (14.83~d) are 
well-established facts. We argue for a $\beta$\,Cep pulsation against HADS and
against a simple eclipsing binary system, a hybrid $\beta$ Cep/SPB star and a 
Be star. The hypothetical solution we propose is either a $\beta$\,Cep star surrounded by a 
corotating dust disk inclined to the equatorial plane or a $\beta$\,Pictoris 
or UXOR star with a regular short-period variability.

These findings make these observations of CoRoT\,102781750 an interesting
astrophysical laboratory.  Our investigation shows the potential of the CoRoT
contaminated light curves and colours.

Further investigation of this star with large telescopes is highly
recommended, as it will serve the interest of multiple fields of astronomy:
active stars, asteroseismology, binary systems, or stellar formation.

\begin{acknowledgements}
MP, JMB, and RSz acknowledge the support of the ESA PECS projects 
No.~98022 \& 98114 and the Hungarian OTKA grant K83790. MP thanks to 
B.~Szeidl, K.~Ol\'ah, K.~Vida, P.~\'Abrah\'am and 
A.~Mo\'or for fruitful discussions. 
KK and EG are supported by the Austrian FWF (Fonds zur F\"o¶rderung der
wissenschaftlichen Forschung) projects T359 and P19962, respectively. 
This research has made use of the Exo-Dat database, operated at LAM-OAMP, 
Marseille, France, on behalf of the CoRoT/Exoplanet programme. 
\end{acknowledgements}

\bibliographystyle{aa}

 
\end{document}